\documentclass{SciPost}
\hypersetup{
    colorlinks,
    linkcolor={red!50!black},
    citecolor={blue!50!black},
    urlcolor={blue!80!black}
}
\usepackage[bitstream-charter]{mathdesign}
\usepackage[utf8]{inputenc}
\usepackage[T1]{fontenc}
\usepackage{ulem}

\newcommand{\vor}[1]{{\color{blue}#1}}

\DeclareSymbolFont{usualmathcal}{OMS}{cmsy}{m}{n}
\DeclareSymbolFontAlphabet{\mathcal}{usualmathcal}

\fancypagestyle{SPstyle}{
\fancyhf{}
\lhead{\colorbox{scipostblue}{\bf \color{white} ~SciPost Physics }}
\rhead{{\bf \color{scipostdeepblue} ~Submission }}

\fancyfoot[C]{\textbf{\thepage}}
}

\begin{document}

\pagestyle{SPstyle}

\begin{center}{\Large \textbf{\color{scipostdeepblue}{
Universal relations in Bose gases with power-law interactions in two and three dimensions
}}}\end{center}

\begin{center}\textbf{
Nina S. Voronova\textsuperscript{1,2$\star$},
Elena V. Skirdova\textsuperscript{1},
I. L. Kurbakov\textsuperscript{3} and
Grigory E. Astrakharchik\textsuperscript{4$\dagger$}
}
\end{center}

\begin{center}
{\bf 1} National Research Nuclear University MEPhI (Moscow Engineering Physics Institute), Moscow, Russia\\
{\bf 2} Russian Quantum Center, Skolkovo Innovation Center, Moscow, Russia\\
{\bf 3} Institute of Spectroscopy of the Russian Academy of Sciences, Troitsk, Moscow, Russia\\
{\bf 4} Departament de Física, Universitat Politècnica de Catalunya, Barcelona, Spain
\\[\baselineskip]
$\star$ \href{mailto:email1}{\small nsvoronova@mephi.ru}\,,\quad
$\dagger$ \href{mailto:email2}{\small grigori.astrakharchik@upc.edu}
\end{center}

\section*{\color{scipostdeepblue}{Abstract}}
\boldmath\textbf{
We derive universal relations between thermodynamic quantities and the pair distribution function for bosons interacting via repulsive power-law potentials in the dilute regime in three and two dimensions. In particular, we obtain an explicit relation between the potential energy and Tan's contact, for which we also find beyond-mean-field expressions, and establish new connections between the chemical potential and the potential energy per particle, characteristic of systems with finite-range interactions. We perform quantum Monte Carlo simulations in order to verify up to what extent the relation between Tan's contact and the chemical potential holds in three- and two-dimensional systems. The obtained results are relevant for dipoles, quadrupoles, Rydberg atoms, excitons, and other systems in the dilute limit. 
}

\vspace{\baselineskip}


\vspace{10pt}
\noindent\rule{\textwidth}{1pt}
\tableofcontents
\noindent\rule{\textwidth}{1pt}
\vspace{10pt}

\section{Introduction}
\label{sec:intro}

Ultradilute atoms provide an excellent platform for high-precision experiments due to their purity and the high degree of controllability of both interactions and external potentials. Quantum gases have enabled experimental observation of a great variety of phenomena such as the BCS-BEC crossover~\cite{Zwerger2012}, superfluid --- Mott Insulator phase transitions~\cite{BlochZweger2008}, formation of solitons~\cite{Kartashov2019} among others, with an unprecedented degree of control not achievable in conventional condensed matter systems. 
At the same time, the majority of the observed phenomena have previously been present in other, perhaps not so clean, systems, e.g. for solid-state quasiparticles such as excitons, exciton-polaritons, etc.~\cite{polaritonBEC-BSC,excitonMott,polariton_solitons}. 
In that sense, the concept of the \textit{contact} stands out as being genuine to ultracold atoms with short-range interactions~\cite{Olshanii,Tan1}. 
It provides a powerful set of relations between thermodynamic quantities and the ultraviolet behavior of the correlation functions (pair distribution function, momentum distribution, one-body density matrix, spectral functions, etc).

The search for universal relations in ultracold gases dates back to 2003, when Maxim Olshanii~\cite{Olshanii} established a relation between the coefficient in the $1/p^4$ high-momentum tail of the momentum distribution, the short-range expansion of the one-body density matrix, and the equation of state of one-dimensional bosons. 
It was soon realized that similar relations hold for fermions~\cite{viverit}. 
The research has culminated in an influential series of works by Shina Tan where the most celebrated relations in terms of `Tan's contact' $\mathcal{C}$~\cite{Tan1,Tan2,Tan3} or a single dimensionless function of the $s$-wave scattering length $a$ and temperature~\cite{ZhangLeggett} have been derived, expressing important thermodynamic observables in terms of the same parameter. 
The contact and the corresponding Tan's relations are exact for zero-range contact interactions and have been theoretically studied~\cite{Braaten08,Combescot09,Braaten10,Braaten11,Matveeva16,Lang17,DeRosi19,Rakhimov20,Bouchoule21,DeRosi21,DeRosi23,Cherny23,Werner24,DeRosi24,Kerr24} and experimentally observed in three-dimensional~\cite{Wild2012,Jin14bec,Clement16,Hadzibabic17} and planar~\cite{Zou2021} Bose gases as well as in unitary Fermi gases and across BCS-BEC crossover~\cite{Jin10,Jin12,Vale19,Zwierlein19,Jager2024}. 
The power of Tan's contact is that it relates the correlation functions to the equation of state. 
The equations of state have a long history of studies and are known both for spin-1/2 fermions in three- (3D)~\cite{Lee1957a,Huang1957}, two- (2D)~\cite{Bloom}, and one-dimensional~\cite{Gaudin} (1D) geometries, as well as for bosons in 3D~\cite{Lee1957b}, 2D~\cite{prl102180404} and 1D~\cite{LiebLiniger63} cases. 

It is a notable feature of ultracold atoms that the range of the interaction potential is significantly smaller than the mean interparticle distance. Under such conditions, the specific form of the interaction potential $V(r)$ is no longer important, and it can be characterized by a single parameter, namely the $s$-wave scattering length $a$. In the dilute regime, where the atomic density $n$ is low, many system properties are universal, i.e. they depend only on the gas parameter $na^3$ (or $na^2$ in the 2D case). In the universal regime, the range of the potential and the effective range in the scattering problem are not important. 
For the contact interactions in three dimensions, Tan obtained a generic relation between the contact and the equation of state~\cite{Tan3}, 
\begin{equation}\label{sweep3D}
\mathcal{C}_{\rm adiab}^{3D} = \frac{8\pi m}{\hbar^2}  \frac{\partial (E/V)}{\partial(-1/a)},
\end{equation}
which was later generalized for two dimensions (see e.g.~\cite{WernerCastinI,WernerCastinII}) as
\begin{equation}\label{sweep2D}
\mathcal{C}_{\rm adiab}^{2D} = \frac{4\pi m}{\hbar^2}  \frac{\partial (E/S)}{\partial(\text{ln}\,a)}.
\end{equation}
Here, $m$ is the particle mass, while $V$ and $S$ denote the system volume (3D) and area (2D), respectively.

The contact calculated according to Eqs.~(\ref{sweep3D}-\ref{sweep2D}) can also be conveniently expressed in terms of the energy per particle $E/N$ and the chemical potential $\mu = \partial E/\partial N$,
\begin{subequations}\label{Eq:Tan:adiabatic}
\begin{equation}\label{Eq:Tan:adiabatic:3D}
\mathcal{C}_{\rm adiab}^{3D} = \frac{8\pi m}{\hbar^2}\;na\left(3\mu - 5\frac{E}{N}\right),
\end{equation}
\begin{equation}\label{Eq:Tan:adiabatic:2D}
\mathcal{C}_{\rm adiab}^{2D} = \frac{8\pi m}{\hbar^2}\;n\left(\mu - 2\frac{E}{N}\right).
\end{equation}
\end{subequations}
The derivation of Eqs.~\eqref{Eq:Tan:adiabatic} from the equations of state is provided in Appendix~\ref{app_sweep}. 

The contact $\mathcal{C}$ is related to the ultraviolet (UV) behavior of the many-body wave function and, according to the adiabatic sweep theorem~(\ref{sweep3D}-\ref{sweep2D}), emerges as a thermodynamic parameter characterizing the variation of the free energy $\mathcal{F}$ with respect to the coupling constant. 
The physical reason for the existence of the contact is that for a zero-range pseudopotential, $V(r) = g \delta({\bf r})$, 
the order of the thermal average $\langle\cdots\rangle$ and the differentiation of the Hamiltonian $H$ with respect to $g$ can be interchanged. This yields either the derivative of a thermodynamic quantity, i.e., the free energy $d\langle H \rangle / dg = d\mathcal{F}/dg$, or the local value of the density-density correlation function $\langle dH/d{\rm g} \rangle \propto \langle {\rm g} \delta({\bf r}) \rangle \propto \langle \delta({\bf r}) \rangle \propto g_2(0)$. 
In this way, the short-range behavior of correlation functions is related to the derivatives of the equation of state with respect to ${\rm g}$ (for an extensive review, see Refs.~\cite{WernerCastinI,WernerCastinII}).
This conceptually simple reasoning is applicable to this regime where the coupling constant $g$ alone is sufficient to describe the interaction, making Tan's relations exact for a contact pseudopotential. 
However, there has been significant experimental progress in realizing ultracold systems with finite-range potentials, such as dipolar gases\cite{Chomaz2022}, Rydberg atoms\cite{Browaeys2020}, atom-ion systems\cite{Harter2014}, where the original description in terms of Tan's contact might no longer be applicable. The generalization of universal relations to finite-range interactions is an open question which is addressed in the present work. Here, we demonstrate that zero-temperature Bose gases with finite-range power-law interactions can be treated outside of the universal regime, i.e. without replacing the real interaction potential with the contact pseudopotential.

We consider a very general model of Bose gases interacting via an arbitrary power-law repulsive potential
\begin{equation}\label{power_potential}
V(r)=\frac{Q}{r^\alpha},
\end{equation}
where $Q$ denotes the amplitude and $\alpha$ the exponent of the two-body interaction potential. 
Additionally, we assume that $V(r)$ is integrable at infinity, i.e. $\alpha>2$ in 2D and $\alpha>3$ in 3D, which ensures that at the many-body level, the interaction~(\ref{power_potential}) decays sufficiently fast to give rise to proper thermodynamics with a finite value of potential energy per particle in the macroscopic limit.  
In the strict sense we consider short-range potentials, although it is common in the cold-gases community to refer to potentials~(\ref{power_potential}) as being long-range ones, in order to differentiate them from zero-range potentials. 
Our aim is to calculate Tan’s contact $\mathcal{C}$ for an arbitrary exponent $\alpha$ and derive universal relations without relying on the adiabatic sweep theorem. 
We perform Monte Carlo simulations in both 2D and 3D geometries to check the agreement between Tan's contact and the adiabatic relation to the equation of state~(\ref{sweep3D}-\ref{sweep2D}) which can be expected to hold when the interaction exponent $\alpha$ is increased to infinity. 
In the following, we address an open question of whether the relations~\eqref{Eq:Tan:adiabatic} hold for power-law potentials, by comparing the contact $\mathcal{C}$ as it comes out of the Monte Carlo pair correlation function with the prediction of Eq.~(\ref{Eq:Tan:adiabatic}) using numerically obtained chemical potential $\mu$ and energy per particle $E/N$.  

In the following, we consider $N$ bosons with coordinates ${\bf r}_1,\dots, {\bf r}_N$. 
When the relative distance between any $i$-th and $j$-th particles is small compared to the mean interparticle separation, 
the influence of all other particles is reduced and the many-body wave function $\Psi({\bf r}_1,\dots,{\bf r}_i,\dots,{\bf r}_j,\dots,{\bf r}_N)$ can be approximately factorized as
\begin{equation}\label{scattering_sep}
\lim\limits_{{\bf r}_i\to{\bf r}_j}\Psi = \psi(r)\tilde{\Psi}({\bf r}_1,\dots,{\bf r}_{i-1},{\bf r}_{i+1}\dots,{\bf r}_{j-1},{\bf r}_{j+1},\dots,{\bf r}_N),
\end{equation}
where $\psi(r)$, ${\bf r}={\bf r}_i-{\bf r}_j$, is the solution for the two-body scattering problem of the relative motion.
The scale at which two adjacent particles ``forget'' about the rest of the gas is defined by $0<r\ll\xi$, where $\xi$ is the healing length (while for the 3D case this is an established fact, see e.g. Ref.~\cite{holzmann_castin}, in order to verify this in the 2D geometry, we have performed variational Monte Carlo simulations that are presented in Appendix~\ref{AppMC}).
On the other hand, on large scales ($r\gg a$, or for $a\to0$), one can describe the collective behavior of the $N$-particle Bose gas and determine the pair distribution function from the Bogoliubov theory\cite{Bogolubov1947} at $T=0$. 
Note that in this reasoning, in Eq.~(\ref{scattering_sep}) one solves the scattering problem on the full interaction potential~(\ref{power_potential}), i.e. without introducing the zero-range approximation. Hence there is no requirement $r\gg a$ for the pair distribution function obtained from the solution of the two-body relative motion Schrödinger equation for $\psi(r)$. 
In the region where these two scales overlap, $a \ll r \ll \xi$, two solutions can be matched. 
Doing so allows us to derive Tan's contact $\mathcal{C}$ in an alternative way, which in two dimensions proves to yield a more accurate result as compared to the one derived from the adiabatic sweep theorem, and in three dimensions recovers the known expressions. 

Additionally, since the considered interactions possess a finite range, within our theory we are able to express the potential energy in terms of the contact $\mathcal{C}$. While for arbitrary power-law interactions in two dimensions this has been done by Hofmann and Zwerger~\cite{HofmannZwerger21}, in a 3D geometry this non-trivial result is obtained here for the first time. Furthermore, using the beyond-mean-field equations of state, we derive the thermodynamic universal relation between the potential energy per particle and the chemical potential of the gas at $T=0$. Expressing Tan's contact in terms of the chemical potential in both 2D and 3D, we check the validity of the derived relations by comparison with Monte Carlo simulations for different power-law exponents $\alpha$.

\section{Two dimensions}
\subsection{Two-body scattering problem and universal relation for potential energy}
To describe correctly the short scales we solve the two-body $s$-wave scattering problem, given by the zero-energy Schrödinger equation of the particle relative motion in power-law potential~(\ref{power_potential})
\begin{equation}\label{2D_scat}
\left[-\frac{\hbar^2}{m} \frac{1}{r}\frac{\partial}{\partial \!r}\left( r \frac{\partial}{\partial \!r}\right) +\frac{Q}{r^\alpha}\right] \psi(r)=0,
\end{equation}
where $m/2$ is the reduced mass. 
The angular dependence in the 2D Laplacian and the wave function is omitted as the lowest-energy solution is symmetric. Our aim is to calculate the pair correlation function $g_2(\textbf{r}_i, \textbf{r}_j)$.

\begin{figure}[t!]
\centering
\includegraphics[width=\textwidth]{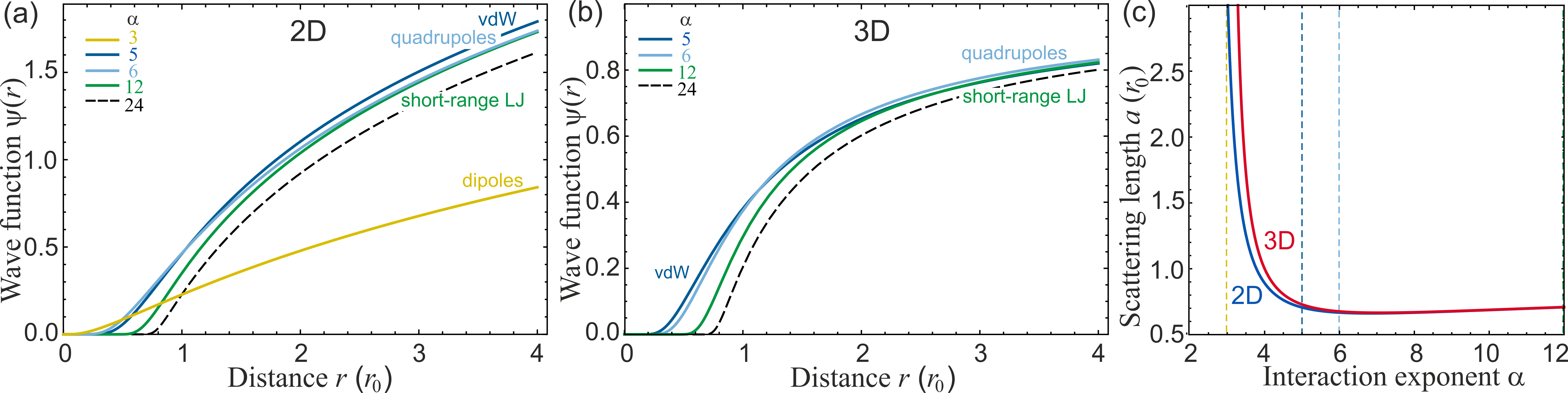}
\caption{\linespread{1.0}\small 
(a,b) Scattering solution $\psi(r)$ in units of the characteristic length $r_0$ associated with interaction potentials.
(a) Two-dimensional scattering problem~(\ref{wavefunc2D}) for different interaction exponents, as marked: $\alpha = 3$ (dipole-dipole interaction), $5$ (quadrupole interaction), $6$ (repulsive tail of van der Waals interaction), $12$ (Lennard-Jones short-range asymptotic). The black dashed line shows the wave function for $\alpha=24$. 
(b) Three-dimensional scattering problem~(\ref{psi3D}). 
Note that while in 2D geometry the solution exhibits a logarithmic divergence at large distances, in the 3D case it remains finite and approaches unity at $r\gg a$ as $1-a/r$. 
(c) The scattering length versus $a$ in units of the characteristic potential range $r_0$ as a function of the power-law exponent $\alpha$ in two- (the blue line, Eq.~(\ref{a_2D})) and three-dimensional cases (the red line, Eq.~(\ref{a_3D})). Vertical dashed lines show guides for the eye for the powers used in the main panels.}
\label{fig1}
\end{figure}

To make the derivation simpler, it is convenient to introduce the dimensionless variables $\rho = r/r_0$ by using as a unit of length the characteristic length associated with the interaction potential, $r_0 = (mQ/\hbar^2)^{\nu}$ with $\nu=1/(\alpha-2)$.
By doing so, Eq.~(\ref{2D_scat}) is reduced to
\begin{equation}\label{eq22}
\psi^{\prime\prime}(\rho) +\frac{1}{\rho}\psi^\prime(\rho) - \frac{1}{\rho^{2+1/\nu}} \psi(\rho)=0,
\end{equation}
which with another substitution $z = 2\nu/\rho^{1/(2\nu)}$ becomes the equation for cylindrical functions, with the general solution $\psi = \ {C_1} I_0(z) +\ {C_2} K_0(z)$, where $I_0(z)$ and $K_0(z)$ are modified Bessel functions. 
We set $C_1=0$ to eliminate the divergent term $I_0(z\to\infty)\to\infty$. 
We then use the asymptotic form $K_0(z\to0)\approx \ln\left(2e^{-\gamma_E}/z\right)$, where $\gamma_E\approx0.577216$ is Euler's constant, to match the obtained solution with the logarithmic large-distance behavior of the wave function $\lim\limits_{r\to\infty}\psi(r)=\ln(r/a)$, which is a solution to the free particle equation $\Delta\psi(r)=0$ in two dimensions.
Thus we define $C_2=2\nu$ and obtain the following expression for the $s$-wave scattering length for scattering on the potential~(\ref{power_potential}):
\begin{equation}\label{a_2D}
a = r_0 \left(\nu e^{\gamma_E}\right)^{2\nu} \!\! = \left[\frac{mQe^{2\gamma_E}}{\hbar^2(\alpha-2)^2}\right]^{1/(\alpha-2)}
\end{equation}
for any $\alpha>2$. 
We note that the same result was given in Ref.~\cite{HofmannZwerger21}. 
For a purely dipolar interaction, $V_d(r)=d^2/r^3$, with dipoles oriented perpendicularly to the plane of motion, this general expression yields $a_d = e^{2\gamma_E} md^2/\hbar^2$ so that the $s$-wave scattering length is proportional to the dipolar length~\cite{ticknor}. 
The scattering solution for potential~(\ref{power_potential}) takes the form
\begin{equation}
\label{wavefunc2D}
\psi(r) = 2\nu K_0\left[2\nu\left(\frac{r_0}{r}\right)^{1/(2\nu)} \right]\quad\text{with}\quad \nu=\frac{1}{\alpha-2}.
\end{equation}
Characteristic examples of the scattering solution~(\ref{wavefunc2D}) and the scattering length~(\ref{a_2D}) are shown in Figs.~\ref{fig1}{\bf a} and {\bf c}, respectively. 
The wave function is significantly suppressed for $r \lesssim r_0$ in the case of the steep potentials ($\alpha>3$), such as those corresponding to quadrupoles, Rydberg atoms with a van der Waals long-range tail, and excitons described by the Lennard-Jones potential. 
The extent of this suppression region (i.e. an effective core radius) increases with the steepness of the potential (i.e., with larger values of $\alpha$), ultimately tending to the hard core of the radius $r_0$ as $\alpha\to\infty$.
Dipolar interactions, $\alpha=3$, rather have a ``soft core'' as the particles are allowed to approach each other more closely.
We note that, considering power-law interactions of an arbitrary exponent $\alpha$, in our theory we do not introduce a short-distance cutoff, as 
was done in Ref.~\cite{HofmannZwerger21} for the particular case of dipolar gases in the broader framework of more realistic interaction potentials deviating from the power-law at small $r$.

The pair distribution function $g_2({\bf r}_i,{\bf r}_j)$ in a homogeneous system depends only on the relative distance $r=|{\bf r}_i - {\bf r}_j|$ and, according to the short-distance behavior of the solution~(\ref{scattering_sep}), it can be universally defined via the two-body wave function (\ref{wavefunc2D}) as
\begin{equation}
\label{g2_2D}
g_2(r) = \frac{\mathcal{C}}{4\pi^2} \psi^2(r),
\end{equation}
where $\mathcal{C}$ is Tan's contact~\cite{WernerCastinII,cherny_2d}. The homogeneity of the system implies $\lim\limits_{r\to\infty}g_2(r) = n^2$, where $n=N/S$ is the 2D particle density with $N$ being the number of particles. In our convention, the two-body scattering solution $\psi(r)$ is dimensionless, while Tan's contact $\mathcal{C}$ and 
the pair distribution function $g_2(r)$ have units of $n^2$. 
Since we consider purely power-law interaction potentials, no other contact arises in our description from the change of the energy with respect to variation of $Q$. This can be contrasted with the case considered in Ref.~\cite{HofmannZwerger21}  where the interaction potential contained both short-range and dipolar terms, leading to the appearance of two contacts.

The relation~(\ref{g2_2D}) between the scattering solution of the two-body problem and the pair correlation function holds in many-body systems but only at short distances\footnote{When any two particles come close to each other, they are weakly affected by the rest of the system and at small distances the pair distribution function $g_2(r)$ can be defined from the scattering problem. As such, if there are just two particles in the box, one has $g_2(r) = 2\psi^2(r)$, where $\psi(r)$ is the normalized wave function of the relative motion. If the box contains $N$ particles, the behavior $g_2(r)\propto\psi^2(r)$ at small $r$ still holds, while the normalization of $\psi(r)$ changes, as the scattering problem is a valid approximation only at distances much smaller than the healing length (see Appendix~\ref{AppMC}). The corresponding proportionality coefficient is $\mathcal{C}/4\pi^2$.}, $r\ll\xi$, where $\xi = \hbar/(\sqrt{2}mc)$ is the healing length and $c$ is the velocity of sound defined at $T=0$ via the inverse compressibility as $mc^2 = n\partial\mu/\partial n$ ($\mu$ is the chemical potential, see e.g.~\cite{PitStr}). Thus, this result, due to the integrability of the interaction potential~(\ref{power_potential}) at infinity, can be used to define the potential energy density of the Bose gas in the limit of small densities,
\begin{equation}
u_{\rm 2D} \equiv \frac{U}{S} = \frac{1}{2} \int \frac{Q}{r^\alpha} g_2(r) d^2\textbf{r} = \frac{\mathcal{C}}{4\pi }\int\limits_0^\infty\frac{Q}{r^\alpha}\psi^2(r)rdr,
\end{equation}
where $U$ is the mean potential energy and the scattering solution $\psi(r)$ is given by~(\ref{wavefunc2D}). Integration yields:
\begin{equation}
\label{U/S}
u_{\rm 2D} = \frac{\hbar^2 \mathcal{C}}{4\pi m(\alpha - 2)},
\end{equation}
which results in a universal relation between the potential energy density and Tan's contact $\mathcal{C}$ valid for power-law interactions with $\alpha>2$ in two dimensions and at any temperature (as no assumption about the temperature has been made up to this point). Relation~(\ref{U/S}) has also been obtained in Ref.~\cite{HofmannZwerger21}. 

Below, we match the short-distance part of the pair correlation function~(\ref{g2_2D}), described by the two-body problem for $0<r\ll \xi$, to the pair correlation function $g_2(r)$ of the full many-body system, obtained within Bogoliubov theory~\cite{Bogolubov1947} and valid for $r\gg a$.
The range of validity of this approach is $a\ll r\ll\xi$.

\subsection{Pair correlation function of the Bose gas from the Bogoliubov theory}

In the second quantization approach, the pair correlation function can be expressed in terms of the bosonic field operators according to $g_2(\textbf{r},\textbf{r}^\prime) = \langle \hat{\Psi}^\dag(\textbf{r}) \hat{\Psi}^\dag(\textbf{r}^\prime) \hat{\Psi}(\textbf{r}^\prime) \hat{\Psi}(\textbf{r}) \rangle$, where the averaging is performed over the ground state of the system at $T = 0$. In the homogeneous case, separating the condensate mode ${\bf p}=0$ in the field operator and performing the Bogoliubov transformation for the non-condensed part
\begin{equation}
\label{Psi_Bog}
\hat{\Psi}(\textbf{r}) = \sqrt{n_0} + \frac{1}{\sqrt{S}} \sum_{\textbf{p} \neq 0 } \ e^{\frac{i\textbf{p}\cdot\textbf{r}}{\hbar}} \big( u_{\textbf{p}} \ \hat{\alpha}_{\textbf{p}} - \ {v}_{-\textbf{p}}^{*} \ {{\hat{\alpha}}^{+}}_{-\textbf{p}} \big),
\end{equation}
provides for the pair correlation function
\begin{equation}
\label{g2_Bog}
g_2(r) =  n_0^2 + \frac{n_0}{S} \bigg[\sum_{\textbf{p} \neq 0} e^{-\frac{i\textbf{p}\cdot\textbf{r}}{\hbar}} \Big( |u_{\textbf{p}} - v_{\textbf{p}}|^2  - 1 \Big) +  2\sum_{\textbf{p} \neq 0} |v_{\textbf{p}}|^2  \bigg].
\end{equation}
In Eqs.~(\ref{Psi_Bog}-\ref{g2_Bog}), $n_0$ is the condensate density, and $\hat{\alpha}$, $\hat{\alpha}^\dag$ denote the Bogoliubov quasiparticle operators obeying the bosonic commutation algebra. The Bogoliubov amplitudes $u_{\bf p}$, $v_{\bf p}$ are given by
\begin{equation}\label{upvp}
\{u_{\bf p}^2, v_{\bf p}^2\} = \frac12\left(\sqrt{1+\frac{(mc^2)^2}{E_p^2}} \pm 1\right),
\end{equation}
with $E_p = \sqrt{p^2/2m(p^2/2m + 2mc^2)}$ being the Bogoliubov excitation spectrum~\cite{AGD}.
Using $(1/S)\sum_{{\bf p}\neq0}|v_{\bf p}|^2 = n-n_0$ (at $T=0$) and changing from the summation to integration over the momenta, with the substitution of (\ref{upvp}) we get from Eq.~(\ref{g2_Bog})
\begin{equation}\label{g2_J0}
\frac{g_2(r)}{n^2} =  1 + \frac{n_0}{n} \int\limits_0^\infty\frac{pdp}{2\pi\hbar^2n}\, J_0\left( \frac{pr}{\hbar}\right)\left(\frac{1}{\sqrt{1+(2mc/p)^2}} - 1\right),
\end{equation}
where $J_0(x)$ is the zero-order Bessel function of the first kind, and the term $\sim (n-n_0)^2/n^2$ was neglected. 
Alternatively, the pair distribution function $g_2(r)$ can be obtained from the Fourier transform of the static structure factor $S(p)$. 
In the single-mode approximation $S(p)$ is related to the Bogoliubov spectrum $E_p$ via the Feynman formula $S(p) = p^2/(2mE_p)$~\cite{feynman,pitaevskii}. 
The resulting expression differs from Eq.~(\ref{g2_J0}) by the absence of the $n_0/n$ factor. This discrepancy arises because, within the Bogoliubov theory up to the considered order of accuracy, there is no distinction between the condensed density $n_0$ and the total density $n$. Therefore, the presence of $n_0$ is an artifact of the Bogoliubov approximation and $n_0$ should be substituted by $n$.

Expression~(\ref{g2_J0}) for the pair correlation function is valid in the long-wavelength limit, i.e. for all $r\gg a$. One of the distinct regimes within this limit is the hydrodynamic range of small momenta ($r\gg\xi$) where $g_2(r) \approx 1- \xi/(2\pi\sqrt{2}nr^3)$. This range lies outside of the universal regime and cannot be matched with the two-body scattering problem. We are interested in the other limit, i.e., the intermediate regime $r\ll\xi$ in which the integration yields
\begin{equation}
\label{g2_Bog_final}
\frac{g_2(r)}{n^2} =  1 - \frac{m^2c^2}{2\pi\hbar^2n} \ln\left(\frac{8}{e^{2\gamma_E+1}}\frac{\xi^2}{r^2}\right) + \mathcal{O}\left( \frac{r^2}{\xi^2}\right).
\end{equation}
The details of this calculation are provided in Appendix~\ref{app_int}.

To match the correlation function~\eqref{g2_Bog_final} with the result of the two-body scattering problem given above, we expand Eq.~(\ref{g2_2D}) with the substitution of Eq.~(\ref{wavefunc2D}) in the limit $r\gg a$ as 
\begin{equation}
\label{g2_log_exp}
\frac{g_2(r)}{n^2} = \mathcal{C} \, \frac{\ln^2(\varkappa \xi^2/a^2)}{(4\pi n)^2} \left[ 1 - 2\frac{\ln(\varkappa \xi^2/r^2)}{\ln(\varkappa \xi^2/a^2)} + \mathcal{O}\left(\ln^{-2}\frac{\varkappa\xi^2}{a^2}\right) \right],
\end{equation}
where we used $\ln(r/a) = [\ln(\varkappa\xi^2/a^2) - \ln(\varkappa\xi^2/r^2)]/2$ and $\varkappa$ is a constant of order 1. While mathematically $\varkappa$ can be arbitrary, our physical goal is to have the matching results obtained in two different approaches. Therefore, comparing Eq.~\eqref{g2_Bog_final} to Eq.~(\ref{g2_log_exp}), we find
\begin{equation}
\label{kappa}
\varkappa = \frac{8}{e^{2\gamma_E+1}} \approx 0.927752.
\end{equation}
Importantly, matching the Eqs.~(\ref{g2_Bog_final}) and (\ref{g2_log_exp}) for $g_2(r)$, which are both valid in the range $a\ll r\ll\xi$, allows us to define Tan's contact $\mathcal{C}$. 

\subsection{Tan's contact and universal relation between the chemical potential and the potential energy density}

We use the beyond-mean-field equation of state of a 2D Bose gas\cite{prl102180404}
\begin{equation}
\label{MoraCastin}
n = \frac{m\mu}{4\pi\hbar^2} \ln{\frac{4\hbar^2}{m\mu a^2 e^{2\gamma_E+1}}}\left(1+\mathcal O\left(\ln^{-2}\frac{\hbar^2\varkappa}{2m\mu a^2}\right)\right).
\end{equation}
In order to get the higher-order relation between the sound velocity $c$ and the $s$-wave scattering length $a$, equivalent to renormalization, we calculate the inverse adiabatic compressibility 
\begin{equation}
\label{g_int}
\frac{\partial\!\mu}{\partial\! n} \approx \frac{4\pi\hbar^2}{m\ln(\hbar^2\varkappa/2m\mu a^2)} + \mathcal O\left(\ln^{-2}\frac{\hbar^2\varkappa}{2m\mu a^2}\right),
\end{equation}
where the last term is a correction, since $a\ll\xi$. 

In terms of the healing length $\xi^2 = \hbar^2/2m^2c^2 =\hbar^2/2m\mu + \mathcal{O}(1)$, therefore
\begin{equation}\label{g_final}
\frac{mc^2}{n} \approx \frac{4\pi\hbar^2}{m\ln(\varkappa\xi^2/a^2)} + \mathcal{O}\left(\ln^{-2}\frac{\varkappa\xi^2}{a^2}\right).
\end{equation}
Substitution into Eq.~(\ref{g2_Bog_final}) yields
\begin{equation}
\frac{g_2(r)}{n^2} =  1 - 2 \frac{\ln(\varkappa\xi^2/r^2)}{\ln(\varkappa\xi^2/a^2)} + \mathcal{O}\left(\ln^{-2}\frac{\varkappa\xi^2}{a^2}\right).
\end{equation}
Comparing to Eq.~(\ref{g2_log_exp}), we find the value of Tan's contact
\begin{equation}
  \label{tan2D}
    \mathcal{C} =   \frac{(4\pi n)^2}{\ln^2(\varkappa\xi^2/a^2)}\left[1 + \mathcal{O}\left(\ln^{-2}\frac{\varkappa\xi^2}{a^2}\right)\right].
\end{equation}
It is important to note that in our approach to the calculation of the contact $\mathcal{C}$ we did not rely on the adiabatic sweep theorem~\cite{Tan2,Tan3,WernerCastinI} (i.e., we did not calculate the variation of the system's energy with the change of the $s$-wave scattering length). Moreover, the result obtained for $\mathcal{C}$ is more accurate compared to the one obtained in the universal regime (see e.g.~\cite{HofmannZwerger21} for dipoles): compared with the previous result, in Eq.~(\ref{tan2D}), the logarithm in the denominator contains $\xi^2$ that is, in turn, also logarithmically dependent on $na^2$, i.e. $\mathcal{C} \sim (n/\ln[\ln(1/na^2)/na^2])^2 + \dots$ [see Eqs.~(\ref{MoraCastin}) and (\ref{g_final})]. The improved accuracy is achieved by matching the two-body scattering problem with Bogoliubov theory by introducing the parameter $\varkappa$. This procedure allows us to determine not only the coefficient under the logarithm but also the subleading correction.
We note that since in Eq.~(\ref{g2_log_exp}) we have used the asymptotics of Eq.~\eqref{wavefunc2D}, i.e. $K_0(z\to0)\simeq \ln(r/a)$, this level of accuracy is not specific to power-law interactions and our derivation applies equally well to short-range potentials. The obtained correction in Eq.~\eqref{tan2D} is, however, specific to the Bessel function and hence to power-law interactions. Furthermore, the interaction exponent $\alpha$ enters the expression for $\mathcal{C}$ via the scattering length.

Finally, expressing the chemical potential $\mu$ from the equation of state~(\ref{MoraCastin})
\begin{equation}\label{chempot}
\mu = \frac{4\pi\hbar^2n}{m\ln(\varkappa\xi^2/a^2)} + \mathcal{O}\left(\ln^{-3}\frac{\varkappa\xi^2}{a^2}\right)
\end{equation}
and comparing with Eq.~\eqref{tan2D}, we obtain a relation similar to Eq.~\eqref{Eq:Tan:adiabatic:2D},
\begin{equation}\label{Ctan_mu_2D}
    \mathcal{C} =  \left(\frac{m}{\hbar^2}\right)^{\!2}\!\mu^2\left[1 + \mathcal{O}\left(\ln^{-2}\frac{\varkappa\xi^2}{a^2}\right)\right].
\end{equation}
Combining with Eq.~(\ref{U/S}), we get the universal relation between the chemical potential and potential energy per particle:
\begin{equation}
\label{uni2D}
\frac{U}{N} = \frac{m\mu^2}{4\pi\hbar^2n(\alpha-2)}
\end{equation}
with a relative error of $\mathcal{O}[\ln^{-2}(\varkappa\xi^2/a^2)]$ at $a\ll\xi$. We note that Eq.~(\ref{uni2D}) depends on the interaction amplitude $Q$ only implicitly.

\begin{figure}[t!]
\centering
\includegraphics[width=\columnwidth]{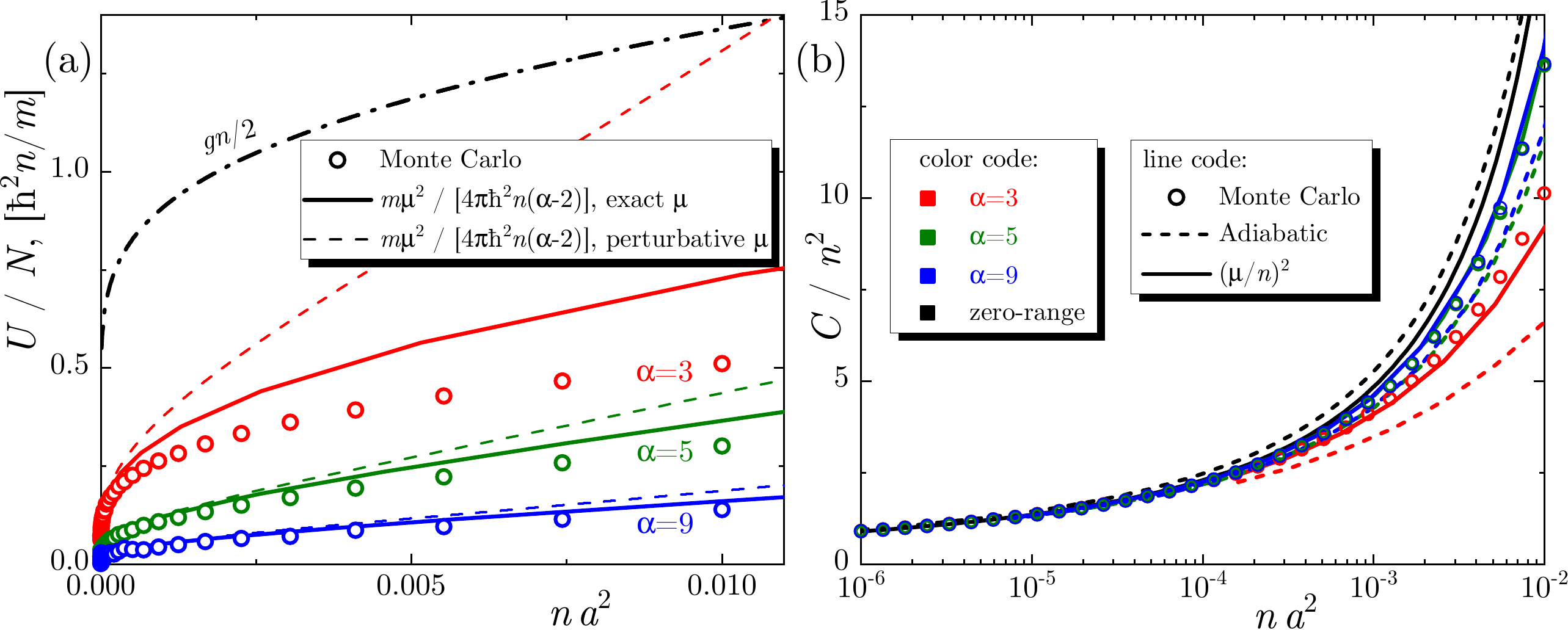}
\caption{\textbf{Properties of systems with power-law interactions in two dimensions.}
(a) Potential energy as a function of the gas parameter for $\alpha = 3; 5; 9$, as obtained in the Monte Carlo simulations (circles) and from Eq.~(\ref{uni2D}) by taking the numerical value of the chemical potential (solid lines) or from a perturbative expansion from Ref.~\cite{PhysRevA79.051602R} (dashed lines).
The dash-dotted line marked with $gn/2$ shows the typical potential energy within Gross-Pitaevskii theory, applied to a zero-range potential.
(b) Tan's contact $\mathcal{C}$ as a function of the gas parameter. Circles show Monte Carlo results extracted by fitting the short-range behavior of the pair distribution function, solid lines of respective colors show the analytical result from Eq.~\eqref{Ctan_mu_2D}. 
}
\label{Fig_CTan2D}
\end{figure}

In Fig.~\ref{Fig_CTan2D}, we compare our analytical results for Tan's contact in terms of the chemical potential squared [see Eq.~\eqref{chempot} and~\eqref{Ctan_mu_2D}] and the potential energy~\eqref{uni2D} with the exact results obtained from Monte Carlo simulations. Details of the simulations and Tan's contact extraction are given in the next Section and Appendix~\ref{app_MC}. This comparison shows a very good agreement of the derived expressions with the {\it ab initio} calculation up to rather large densities. When the potential exponent $\alpha$ is small (i.e. for ``soft'' potentials), the scattering length at zero energy, which was calculated in Eq.~\eqref{a_2D} deviates considerably from the scattering length at the actual energy of colliding particles (of the order of $\mu$)~\cite{pra81013612}. This deviation leads to the observed discrepancy for $\alpha=3$, while for quadrupoles ($\alpha=5$) the agreement is remarkable. 
 
\section{Three dimensions}

For a three-dimensional Bose gas at $T=0$ interacting repulsively via the potential~(\ref{power_potential}), we will develop the theory of Tan's contact and obtain universal relations in analogy with the 2D case. Differently from the two-dimensional case, our considerations here do not include dipolar interactions $\alpha=3$, since the matching of the two-body scattering problem at low energies with the many-body problem in the macroscopic limit is valid only for $\alpha>3$ (i.e., quadrupole interaction, repulsive tail of van der Waals potentials, etc.).

\subsection{Two-body scattering problem and universal relation for potential energy}

We consider the 3D Schrödinger equation of the relative two-body motion at zero energy:
\begin{equation}
\label{Schroedinger3D}
\left[-\frac{\hbar^2}{m} \frac{1}{r^2}\frac{\partial}{\partial\!r}\left(r^2 \frac{\partial}{\partial\!r}\right) +\frac{Q}{r^\alpha} \right] \psi(r)=0
\end{equation}
with the boundary conditions $\psi(0)=0$ and $\psi(\infty)=1$. Using dimensionless units, $\rho = r/r_0$ with $r_0 = (mQ/\hbar^2)^\nu$, $\nu = 1/(\alpha - 2)$ and $z = 2\nu/\rho^{1/(2\nu)}$ (same as before) brings Eq.~(\ref{Schroedinger3D}) to the modified Bessel equation with the general solution 
$$\psi(\rho) = \tilde C_1 \frac{(2\nu)^\nu}{\sqrt{\rho}} I_\nu\left[\frac{2\nu}{\rho^{1/(2\nu)}}\right] + \tilde C_2 \frac{(2\nu)^\nu}{\sqrt{\rho}} K_\nu\left[\frac{2\nu}{\rho^{1/(2\nu)}}\right],$$
where $I_\nu(z)$ and $K_\nu(z)$ are the modified Bessel functions of the $\nu$-th order. 
Similarly to the 2D case, the divergence at $\rho\to0$ is avoided by setting $\tilde C_1=0$.
The $s$-wave scattering length is obtained from the long-range expansion $\psi \sim x^\nu K_\nu(x)$, $K_\nu(x\to0) \approx x^\nu 2^{-1-\nu}\Gamma(-\nu) + x^{-\nu} 2^{-1+\nu} \Gamma(\nu)$ by comparison with the asymptotic solution of the scattering problem $\lim\limits_{r\to\infty} \psi(r) = 1 - a/r$.
The 3D $s$-wave scattering length can be calculated explicitly as
\begin{equation}\label{a_3D}
a = \nu^{2\nu}\frac{\Gamma(1-\nu)}{\Gamma(1+\nu)}r_0 = \frac{\Gamma[(\alpha-3)/(\alpha-2)]}{\Gamma[(\alpha-1)/(\alpha-2)]} \left(\frac{mQ}{\hbar^2(\alpha-2)^2}\right)^{1/(\alpha-2)}\!\!\!,
\end{equation}
where $\Gamma(x)$ is the Gamma function. 
The solution to the two-body scattering problem~(\ref{Schroedinger3D}) reads
\begin{equation}
\label{psi3D}
\psi(r) = \frac{2\nu^\nu}{\Gamma(\nu)} \sqrt{\frac{r_0}{r}} K_\nu \left[2\nu\left(\frac{r_0}{r}\right)^{1/(2\nu)}\right]\quad\text{with}\quad \nu=\frac{1}{\alpha-2}.
\end{equation}
We show characteristic examples of the resulting wave function in Fig.~\ref{fig1}{\bf b}. 
Notably, the ratio of the scattering length in three~(\ref{a_3D}) and two~(\ref{a_2D}) dimensions approaches unity, 
\begin{equation}
\frac{a_{3D}}{a_{2D}}=
\frac{\Gamma(1-\nu)}{e^{2\nu\gamma_E}\Gamma(1+\nu)}\to 1, \quad\alpha\to\infty. 
\end{equation}
The physical reason behind this is that steeper potentials (i.e., those with larger values of $\alpha$) resemble hard cores more closely. For a hard-core potential, the $s$-wave scattering length $a$ is equal to the diameter of the hard core in any dimensionality. Thus Fig.~\ref{fig1}\vor{\bf c} shows the behavior of the scattering length $a$ with the interaction exponent for both 2D and 3D cases, and there is no need to introduce different notations.

As we have defined $\psi(r)$ to be dimensionless 
(that is, its long-range asymptotic value is $\psi(r)\approx 1-a/r$), the relation between the pair correlation function at short distances and the wave function of the two-body scattering problem~\cite{WernerCastinII,cherny_3d} that preservs the correct units is
\begin{equation}
\label{g2_3D_asymp}
\lim\limits_{r\to0} g_2(r) = \frac{\mathcal{C}}{16\pi^2a^2} \psi^2(r).
\end{equation}
For interaction potentials that are short-range in the strict sense (i.e., integrable at infinity), the limiting procedure~(\ref{g2_3D_asymp}) implies that the distance scales are much smaller than the healing length $r\ll\xi$, where $\xi = 1/\sqrt{8\pi an}$ and $n=N/V$ is the 3D particle density.
As the main part of the potential energy comes from the ultraviolet contribution, the mean potential energy can be explicitly calculated relying on Eq.~(\ref{g2_3D_asymp})
\begin{equation}
\label{U/V}
u_{\rm 3D} \equiv \frac{U}{V} = \frac{1}{2} \int \frac{Q}{r^\alpha} g_2(r) d^3\textbf{r} = \frac{\hbar^2\mathcal{C}}{8 \pi a m (\alpha-2)}.
\end{equation}

Combining the large-distance asymptotics of Eq.~(\ref{psi3D}), i.e. $\psi(r) \approx 1 - a/r + \mathcal{O}[(a/r)^{1/\nu}]$, with the definition of the pair distribution function (\ref{g2_3D_asymp}), we arrive at the expression directly suitable for matching with the Bogoliubov expression in the range $a\ll r\ll\xi$:
\begin{equation}\label{g2_3D_stitching}
\frac{g_2(r)}{n^2} = \frac{\mathcal{C}}{16\pi^2a^2n^2}\left[1 - \frac{2a}{r} + \mathcal{O}\left(\frac{a^2}{r^2}\right) + \mathcal{O}\left(\frac{a}{r}\right)^{1/\nu}\right]
\end{equation}
which represents the desired connection between $g_2(r)$ and Tan's contact $\mathcal{C}$ following from the scattering problem.

\subsection{Tan's contact and universal relation in 3D geometry}

For the many-body limit of the pair correlation function within the Bogoliubov theory, after integration over angles in 3D space, we get (cf. Eq.~(\ref{g2_J0})):
\begin{equation}\label{g2_3D_Bog}
  \frac{g_2(r)}{n^2} = 1 - \frac{8a}{\pi r}\int\limits_0^\infty \frac{t^2dt}{16\pi anr^2}\frac{\sin t}{t} \left(1 - \frac{1}{\sqrt{1 + 16\pi anr^2/t^2}}\right),
\end{equation}
where $t = pr/\hbar$ and we used the definition of the coupling constant in 3D ${\rm g} = 4\pi\hbar^2a/m$ (see e.g.~\cite{PitStr}). Given that $16\pi anr^2 = 2(r/\xi)^2\ll1$ in the range where we aim to match with the result of the scattering problem (\ref{g2_3D_stitching}), the integration yields (the details of the calculation are presented in Appendix~\ref{app_int}):
\begin{equation}\label{g2_3D_final}
  \frac{g_2(r)}{n^2} = 1 - \frac{2a}{r} + \frac{64}{3\sqrt{\pi}}\sqrt{na^3} 
  + \mathcal{O}(a^2).
\end{equation}
Comparing to Eq.~(\ref{g2_3D_stitching}), we obtain an explicit expression for Tan's contact
\begin{equation}\label{C_3D}
  \mathcal{C} = (4\pi na)^2\left(1 +  \frac{64}{3\sqrt{\pi}}\sqrt{na^3}+\cdots\right). 
\end{equation}
The obtained expression for the two-body contact for a power-law interacting Bose gas, as well as Eq.~\eqref{g2_3D_final} for the second-order correlation function, coincide with the prediction of the zero-temperature Bogoliubov theory for a contact pseudopotential (see e.g.~\cite{Jin2012, PhysRevA.91.043631}) when including the Lee-Huang-Yang correction~\cite{Lee1957b} to the mean-field equation of state. 

Some additional useful relations can be obtained by using the beyond-mean-field equation of state
\begin{equation}\label{LHY}
\frac{E}{V} = \frac{2\pi\hbar^2an^2}{m}\left(1+\frac{128}{15\sqrt{\pi}} \sqrt{na^3} + \mathcal{O}(na^3)\right),
\end{equation}
which, by differentiating with respect to the density, yields the chemical potential
\begin{equation}\label{mu_3D}
\mu = \frac{\partial}{\partial\!n}\frac{E}{V} = \frac{4\pi\hbar^2an}{m} \left(1 + \frac{32}{3\sqrt{\pi}}\sqrt{na^3} + \mathcal{O}(na^3)\right).
\end{equation}
Comparing with Eq.~\eqref{C_3D}, we recover the following relation (cf. Eq.~\eqref{Eq:Tan:adiabatic:3D})
\begin{equation}\label{Ctan_mu_3D}
    \mathcal{C}= \left(\frac{m}{\hbar^2}\right)^{\!2}\mu^2,
\end{equation}
identical to Eq.~\eqref{Ctan_mu_2D} obtained in two dimensions. Finally, 
combining Eqs.~(\ref{U/V}), (\ref{C_3D}) and (\ref{mu_3D}), we get the universal connection between the chemical potential and potential energy per particle in three dimensions:
\begin{equation}\label{uni3D}
\frac{U}{N} = \frac{m\mu^2}{8\pi\hbar^2an(\alpha-2)}
\end{equation}
with a relative error $\mathcal{O}(na^3)$ at $a\ll\xi$ (which is equivalent to the diluteness condition $na^3\ll1$). We note that Eq.~(\ref{uni3D}) differs from its two-dimensional counterpart Eq.~(\ref{uni2D}) by the dimensional factor of $1/(2a)$.

\begin{figure}[b!]
\centering
\includegraphics[width=0.6\columnwidth]{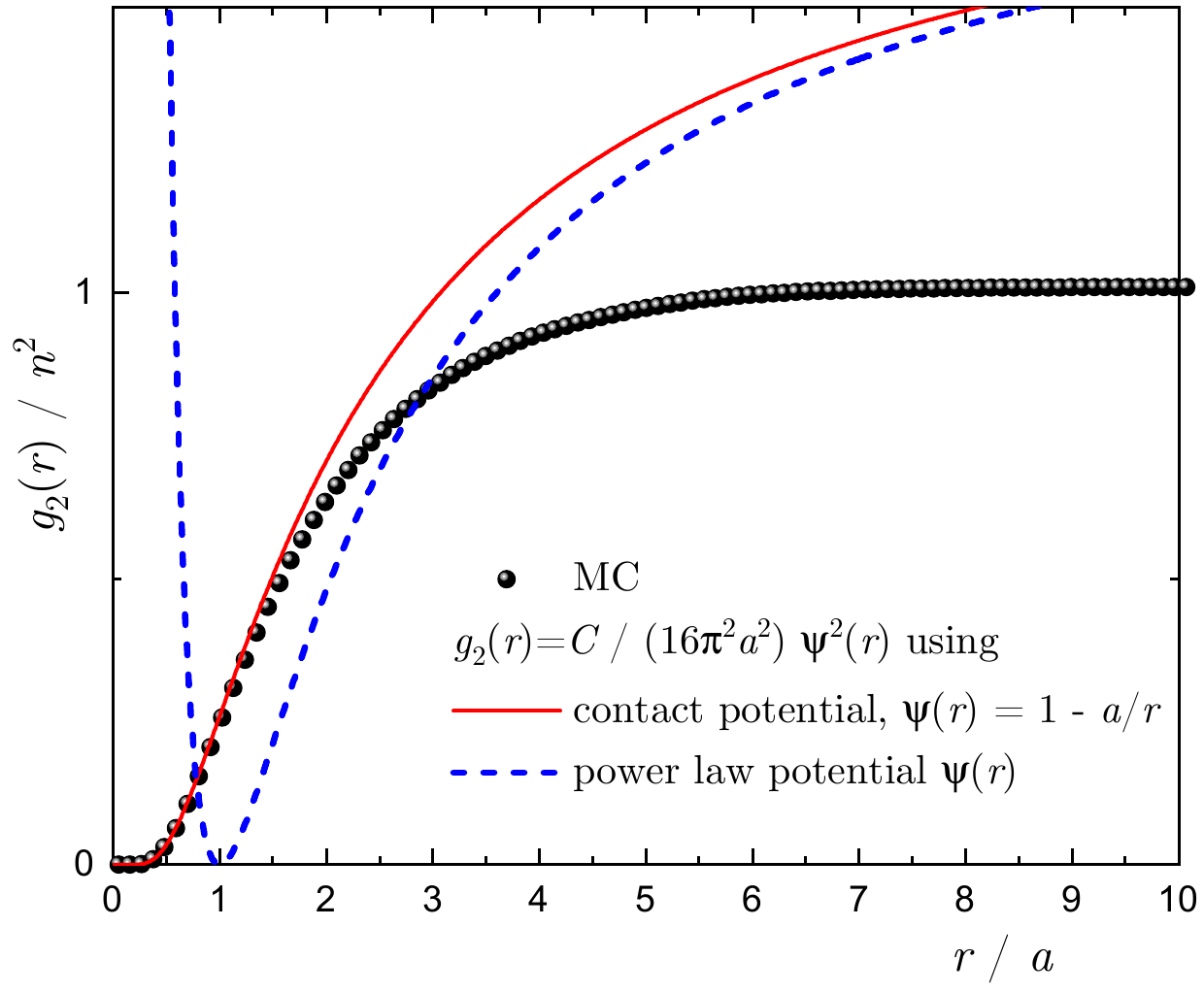}
\caption{\textbf{Characteristic example of the pair distribution function used to extract Tan's contact in a typical 3D Monte Carlo simulation.}
Symbols show the pair distribution function $g_2(r)$ obtained from Monte Carlo simulations. 
Lines represent the short-range asymptotic form $g_2(r)=\mathcal{C}/(16\pi^2a^2)\;\psi^2(r)$, given by Eq.~(\ref{g2_3D_asymp}). The dashed line corresponds to the zero-range limit, $\psi(r)=1-a/r$, consistent with the standard definition of Tan's contact for a zero-range interaction. The solid line uses the two-body scattering solution given by Eq.~(\ref{psi3D}), with $\mathcal{C}$ fitted to reproduce the short-range behavior. 
}
\label{Figg2}
\end{figure}

\begin{figure}[t]
\centering
\includegraphics[width=\columnwidth]{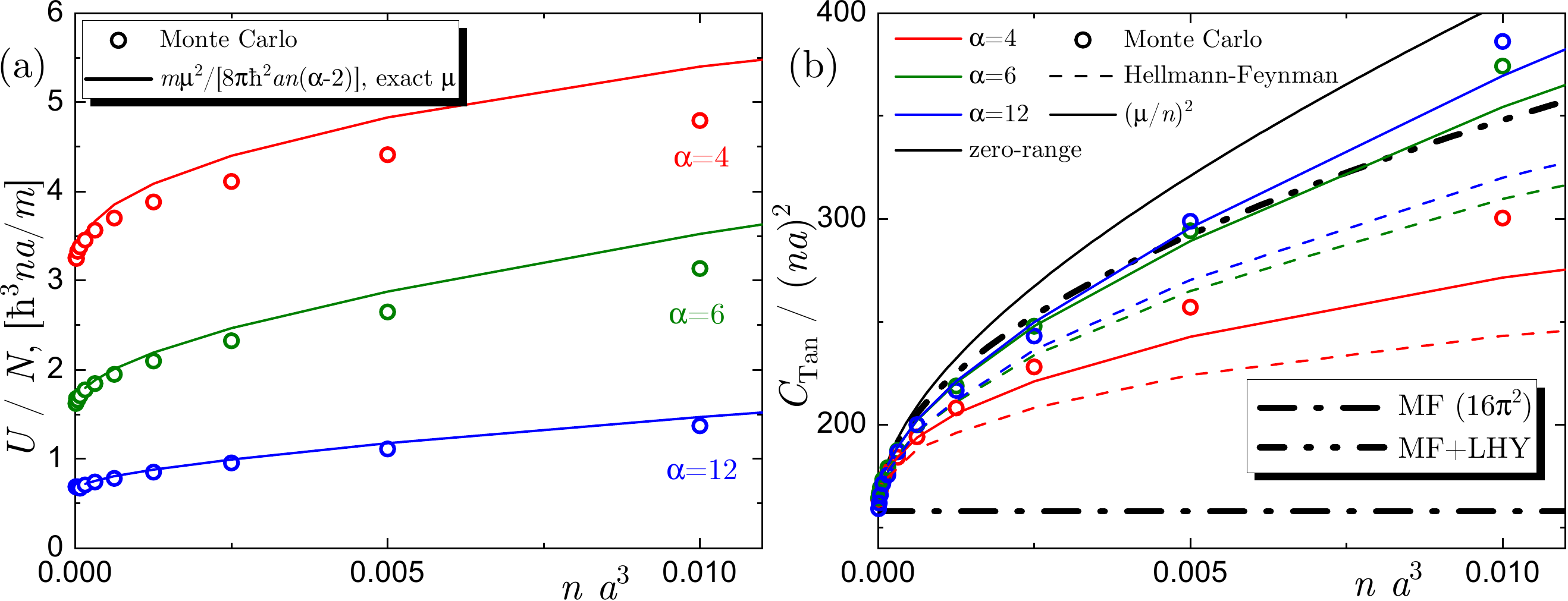}
\caption{\textbf{Properties of systems with power-law interactions in three dimensions.} 
(a) Potential energy per particle, in characteristic units of $\hbar^3 n a/m$, as a function of the gas parameter for interparticle potentials decaying as $r^{-\alpha}$, with $\alpha=4,6,$ and $12$ (coded by different colors, as indicated in the legend). Solid and dashed lines represent Monte Carlo results and predictions from Eq.~(\ref{uni3D}), respectively.
(b) Tan's contact $\mathcal{C}$ as a function of the gas parameter. Circles show the Monte Carlo results extracted from the short-range behavior of the pair distribution function, the solid lines of the respective colors correspond to relation~\eqref{Ctan_mu_3D} to the chemical potential, the black solid line shows the behavior for the zero-range adiabatic-theorem result~\eqref{Eq:Tan:adiabatic:3D}. The dashed, dot-dashed and double-dot dashed lines show various mean-field approximations as marked on the panel.
}
\label{Fig_CTan3D}
\end{figure}

We perform {\it ab initio} Monte Carlo simulations to verify the derived relations (see Appendix~\ref{app_MC} for further details on the method). The contact $\mathcal{C}$ is determined from the short-distance behavior of the pair correlation function $g_2(r)$ using its short-range asymptotics~\eqref{g2_3D_asymp}. A characteristic example is shown in Fig.~\ref{Figg2}. We note that, for a finite-range potential, the actual short-range behavior differs from the zero-range limit. In particular, the zero-range model introduces a node at $r=a$, whereas the ground-state wave function for a finite-range potential is generally nodeless for $r>0$, apart from a possible node at the origin for sufficiently singular power-law potentials.

Figure~\ref{Fig_CTan3D}(a) shows the potential energy calculated numerically and that obtained using relation~\eqref{uni3D} to the chemical potential.  One sees that the perturbative result~\eqref{uni3D} works better for smaller values of the gas parameter $na^3$ and for interactions which decay faster, i.e. those with larger power law $\alpha$. Although the derived expression is not expected to remain quantitatively accurate in the strongly correlated regime, it still provides a useful estimate of the potential energy.
The comparison of various analytical results for Tan's contact $\mathcal{C}$ (our Eq.~\eqref{Ctan_mu_3D}, the expression obtained from the Hellmann-Feynman theorem~\cite{cherny_3d}, the mean-field expressions obtained for the contact pseudopotential, including those with the Lee-Huang-Yang correction) with simulations is summarized in Figure~\ref{Fig_CTan3D}(b). One clearly sees that while for larger exponents $\alpha$ the numerical results, as expected, approach the result of the zero-range pseudopotential, the analytical results obtained in the work for finite $\alpha$ (here, $\alpha = 4, 6, 12$) are in better agreement with simulation results than any other approximation.

\section{Summary}
To conclude, we have derived universal relations for uniform quantum Bose gases with finite-range power-law interactions $V(r) \propto 1/r^\alpha$ in two and three dimensions. Importantly, our results do not rely on the adiabatic sweep theorem, differently from the original approach used by Shina Tan~\cite{Tan2,Tan3}. As a result, the obtained expressions are applicable to potentials that cannot be substituted by zero-range ones. We explicitly define the zero-temperature Tan's contact parameter for any interaction exponent ($\alpha>2$ in 2D and $\alpha>3$ in 3D) without introducing the contact pseudopotential approximation. Instead, we match the pair correlation function found from the exact solution of the low-energy two-body scattering problem (valid for distances smaller than the healing length) and that obtained from the many-body Bogoliubov approximation (valid in the long-wavelength limit). 

In the two-dimensional case, we find a more accurate value for the contact $\mathcal{C}$ than that obtained from the adiabatic sweep theorem (for dipoles, see Ref.~\cite{HofmannZwerger21}). For three dimensions, we show that the contact coincides with that for gases with zero-range interactions, however, in our case, the dependence on $\alpha$ is explicitly included via the scattering length. 
We note that in three dimensions, we find the next-order term with respect to the gas parameter $\propto\sqrt{na^3}$, which does not appear in the previous literature. This beyond-mean-field term in the expression of the contact agrees with the result obtained from the adiabatic theorem.
To check the validity and improved accuracy of our findings compared to the existing results, we performed quantum Monte Carlo simulations. Such a verification allows us to conclude that, in the weak-correlation regime, our theory accurately describes the properties of power-law potentials (i.e., without replacing the interaction potential with $g\delta(r)$). 
In the regime of stronger correlations, the contact extracted from the short-distance behavior of $g_2(r)$ and that obtained from the adiabatic sweep theorem start to deviate, suggesting that no universal contact can be defined for strong correlations. 
We find that defining the contact as the proportionality coefficient relating the pair correlation function $g_2(r)$ to $\psi^2(r)$ of the scattering problem in the limit $r\to 0$, see Eq.~\eqref{g2_2D} (Eq.~\eqref{g2_3D_asymp}) for the 2D (3D) geometry, yields the best agreement with {\it ab initio} simulations when using the universal relations derived here, Eqs.~\eqref{Ctan_mu_2D} and~\eqref{Ctan_mu_3D}. These relations, which predict $\mathcal{C}\propto\mu^2$ in both 2D and 3D geometries, agree remarkably well with Monte Carlo results, unlike the contact obtained from the adiabatic sweep theorem.
The difference between the two definitions becomes more pronounced for more rigid potentials (larger $\alpha$), since for such potentials the dependence of the scattering length on the energy weakens~\cite{pra81013612}.
We note here that another commonly used way to determine the contact is through the high-momentum tail of the momentum distribution, which, in the case of zero-range interactions, decays as $n(\mathbf{p}) \propto \mathcal{C}/p^4$~\cite{Olshanii,cherny_2d,cherny_3d}. Whether the same relation remains in agreement with the definition of $\mathcal{C}$ via $g_2(r)$ for the case of finite-range interactions remains a question to be addressed in future works.

For the case of finite-range interactions considered here, we also define the potential energy density in terms of the contact and therefore show the validity of thermodynamic relations. Including beyond-mean-field corrections to the equation of state (see Mora and Castin~\cite{prl102180404} for the 2D case and Lee, Huang and Young~\cite{Lee1957b} for three dimensions) allows one to accurately connect the chemical potential with the potential energy per particle. Results obtained in this work, including the new universal relations, are summarized in Table~\ref{table1}.

\begin{table}[h]
    \centering
    \begin{tabular}{cc}
    \hline\hline
       \hspace{20pt}{\small Two dimensions}\hspace{20pt}  & \hspace{20pt}{\small Three dimensions}\hspace{20pt} \\
    \hline
    $\alpha > 2$ & $\alpha>3$ \\
    $\displaystyle a = e^{2\gamma_E/(\alpha-2)} \left[\frac{mQ}{\hbar^2(\alpha-2)^2}\right]^{\frac{1}{\alpha-2}} $ & \hspace{10pt} $\displaystyle a = \frac{\Gamma[(\alpha-3)/(\alpha-2)]}{\Gamma[(\alpha-1)/(\alpha-2)]} \left[\frac{mQ}{\hbar^2(\alpha-2)^2}\right]^{\frac{1}{\alpha-2}}$ 
    \\[20pt]
    $\displaystyle \mathcal{C} \!=\!   \frac{(4\pi n)^2}{\ln^2\left[\frac{\ln\{\ln(\dots)/(\pi e^{2\gamma_E + 1} n a^2)\}}{\pi e^{2\gamma_E + 1} na^2}\right]}= \left(\frac{m}{\hbar^2}\right)^{\!2}\!\!\mu^2$ 
    & $\,\,\displaystyle \mathcal{C} = (4\pi na)^{2}\!\left(1 \!+\!  \frac{64}{3\sqrt{\pi}} \sqrt{na^3}\right)= \left(\frac{m}{\hbar^2}\right)^{\!2}\!\!\mu^2$ 
    \\[20pt]
    $\displaystyle \frac{U}{N} = \frac{m}{4\pi\hbar^2n}\frac{\mu^2}{(\alpha-2)}$ &  $\displaystyle \frac{U}{N} = \frac{m}{8\pi\hbar^2an}\frac{\mu^2}{(\alpha-2)}$ \\[15pt]
    \hline\hline
    \end{tabular}
    \caption{The scattering length $a$, Tan's contact $\mathcal{C}$, and the relations involving potential energy $U$ and chemical potential $\mu$ of the zero-temperature Bose gas with power-law interactions, for two and three dimensions. Here, $\alpha$ denotes the power-law exponent of the interaction potential and $Q$ its amplitude. Note that a similar relation between the contact and the chemical potential also holds for contact interactions, as can be seen from Bogoliubov theory.
    }
    \label{table1}
\end{table}

\section*{Acknowledgements}
\paragraph{Funding information}
This work has been supported by the Foundation for the Advancement of Theoretical Physics and Mathematics ``BASIS'' (grant No.~24--1--2--22--1). G.A. is financially supported by the Spanish Ministry of Science and Innovation (MCIN/AEI/10.13039/501100011033, grant PID2023-147469NB-C21)
by the Spanish Ministry of University (grant FPU No. FPU22/03376 funded by MICIU/AEI/10.13039/501100011033), and by the Generalitat de Catalunya (grant 2021 SGR 01411). I.L.K. acknowledges the support of the Project No.~FFUU-2024-0003. 
We acknowledge computational facilities provided by {Barcelona Supercomputing Center MareNostrum} ({FI-2026-1-0076}).

\appendix
\section{The adiabatic-sweep-theorem contact from equations of state} \label{app_sweep}

In this Appendix, we derive relations~\eqref{Eq:Tan:adiabatic} between the contact, the energy per particle and the chemical potential in 3D and 2D geometries.
Since for a large number of particles $N\to\infty$ the only parameter of the problem is the dimensionless particle density $nr_0^d$ (where $r_0$ is the characteristic length scale for the corresponding dimension, ${d} = 2,3$), the energy density can be expressed as
$$\frac{E}{V_{d}} = \frac{\hbar^2}{mr_0^2}\frac{1}{r_0^d}\, f(nr_0^{d})$$
with $V_{d} = N/n$ (i.e. $V_2 \equiv S$, $V_3\equiv V$). Restoring dimensions and expressing lengths in terms of the scattering length $a$, one gets
\begin{equation}
    \frac{E}{V_{d}} = \frac{2\pi\hbar^2n^2 a^{d -2}}{m}\, \mathcal{U}(n a^{d})
\end{equation}
with $\mathcal{U}(A x) = f(x)/x^2$, $A=(a/r_0)^d$. Then the chemical potential
\begin{equation}
    \mu = \frac{\partial E}{\partial N} = \frac{\partial (E/V_d)}{\partial n} = \frac{2\pi\hbar^2 n a^{d-2}}{m}\left[2\,\mathcal{U}(na^d) + na^d\,\mathcal{U}^\prime(na^d)\right].
\end{equation}

\subsection*{Three dimensions}
Using the Lee--Huang--Young equation of state~\eqref{LHY},
\begin{multline}\nonumber
\frac{\partial(E/V)}{\partial a} = \frac{E/V}{a} + \frac{6\pi\hbar^2 n^3 a^3}{m} \, \mathcal{U}^\prime(na^3) = \frac{E/V}{a} + \frac{6\pi\hbar^2 n^2}{m} \left( \frac{\mu m}{2\pi\hbar^2 na} - 2 \frac{m \,E/V}{2\pi \hbar^2 n^2 a} \right) \\
= \frac{E/V}{a} + \frac{3n}{a}\left(\mu - 2\frac EN\right) = \frac{3n}{a} \left(\mu - \frac{5}{3}\frac EN\right),
\end{multline}
so that from the adiabatic sweep theorem Eq.~\eqref{sweep3D} we get:
\begin{equation}
    \mathcal{C}_{\rm adiab}^{3D} = \frac{8\pi m}{\hbar^2} \frac{\partial (E/V)}{\partial (-1/a)} = \frac{8\pi m a^2}{\hbar^2} \frac{\partial (E/V)}{\partial a} = 24\pi na \frac{m}{\hbar^2} \left(\mu - \frac{5}{3}\frac{E}{N}\right).
\end{equation}

\subsection*{Two dimensions}
The same calculation with Mora and Castin's equation of state~\eqref{MoraCastin} yields
$$\frac{\partial(E/S)}{\partial a} = \frac{4\pi \hbar^2 n^3 a}{m} \, \mathcal{U}^\prime(na^2) = \frac{4\pi\hbar^2 n^2}{ma} \left(\frac{\mu m}{2\pi\hbar^2 n} - 2\frac{m\, E/S}{2\pi\hbar^2 n}\right) = \frac{2n}{a}\left(\mu-2\frac EN\right).$$
Substituting into Eq.~\eqref{sweep2D}, we obtain
\begin{equation}
    \mathcal{C}_{\rm adiab}^{2D} = \frac{4\pi m}{\hbar^2}\frac{\partial(E/S)}{\partial(\text{ln}\,a)} = \frac{4\pi ma}{\hbar^2} \frac{\partial(E/S)}{\partial a} = 8\pi n \frac{m}{\hbar^2} \left(\mu-\frac{2E}{N}\right).
\end{equation}

\section{Finding the range of the applicability of the two-body scattering in the many-body problem} \label{AppMC}

We perform variational Monte Carlo simulations in two dimensions for the $V(r)\propto 1/r^6$ interaction potential considered as an example to determine the maximal distances at which the two-body problem remains relevant within a many-body context. 

To do so, we consider the many-body trial wave function in a pair product form of Jastrow terms, constructed from the two-body scattering solution at short distances and a hydrodynamic asymptotic behavior\cite{ReattoChester1967} at large distances:
\begin{equation}\label{wf_trial}
    \Psi_{\rm trial}({\bf r}_1,...{\bf r}_N)=\prod_{1\le j<k\le N}f(|{\bf r}_j-{\bf r}_k|).
\end{equation}
More precisely, the Jastrow factor $f(r)$ is defined as follows: when the two particles approach each other (small $r$), the function $f(r)=C_1\psi_{\rm scat}(r)$ up to a constant $C_1$ approximately coincides with the solution of the scattering problem [see Eq.~\eqref{2D_scat}] with the energy $\mathcal{E}_{\rm par}$. 
At large $r$, the dominant role for the ground state is played by the collective (phonon) behavior. Analytic solution of the two-particle problem in 2D geometry yields~\cite{ReattoChester1967} $f(r)\propto\exp\{-\rm{const}/r\}$. 
For the simulation box of the size $L=(N/n)^{1/2}$, at $r>L/2$ the Jastrow factor is taken constant: $f(r>L/2)=f_2(L/2)$.
To ensure the continuity of both the function and its first derivative at $r=L/2$, the expression for $f(r)$ is symmetrized as $f(r)=C_2\exp\{-C_3/r-C_3/(L-r)\}$. As a result, the function
$f(r)$ and its first derivative $f'(r)$ become continuous both at the box boundary and at $r=L/2$. Furthermore, the function $f(r)$ automatically satisfies periodic boundary conditions. Finally, the normalization condition $f(L/2)=1$ leads to the result
\begin{equation}\label{fr}
f(r)=\left\{\begin{array}{ll}C_1\psi_{\rm scat}(r),\;&r\le R_{\rm par},\\
C_2\exp\{-C_3/r-C_3/(L-r)\},\;&R_{\rm par}\le r\le L/2,\\
1,\;&L/2\le r.\end{array}\right.
\end{equation}
Here $R_{\rm par}$ is the distance where the short-range and long-range solutions are continuously matched. Three continuity conditions, for $f(r)$ at $r=R_{\rm par}$ and $r=L/2$ and for $f'(r)$ at $r=R_{\rm par}$, allow us to unambiguously define the constants $C_1$, $C_2$ and $C_3$. Thus,
the trial wave function~\eqref{wf_trial} is parameterized by the two variables, the scattering energy of the two-body problem 
$\mathcal{E}_{\rm par}$ and the matching distance $R_{\text{par}}$. These parameters are optimized by minimizing the variational energy in Monte Carlo simulations.

\begin{figure}[t!]
\centering
\includegraphics[width=0.7\textwidth]{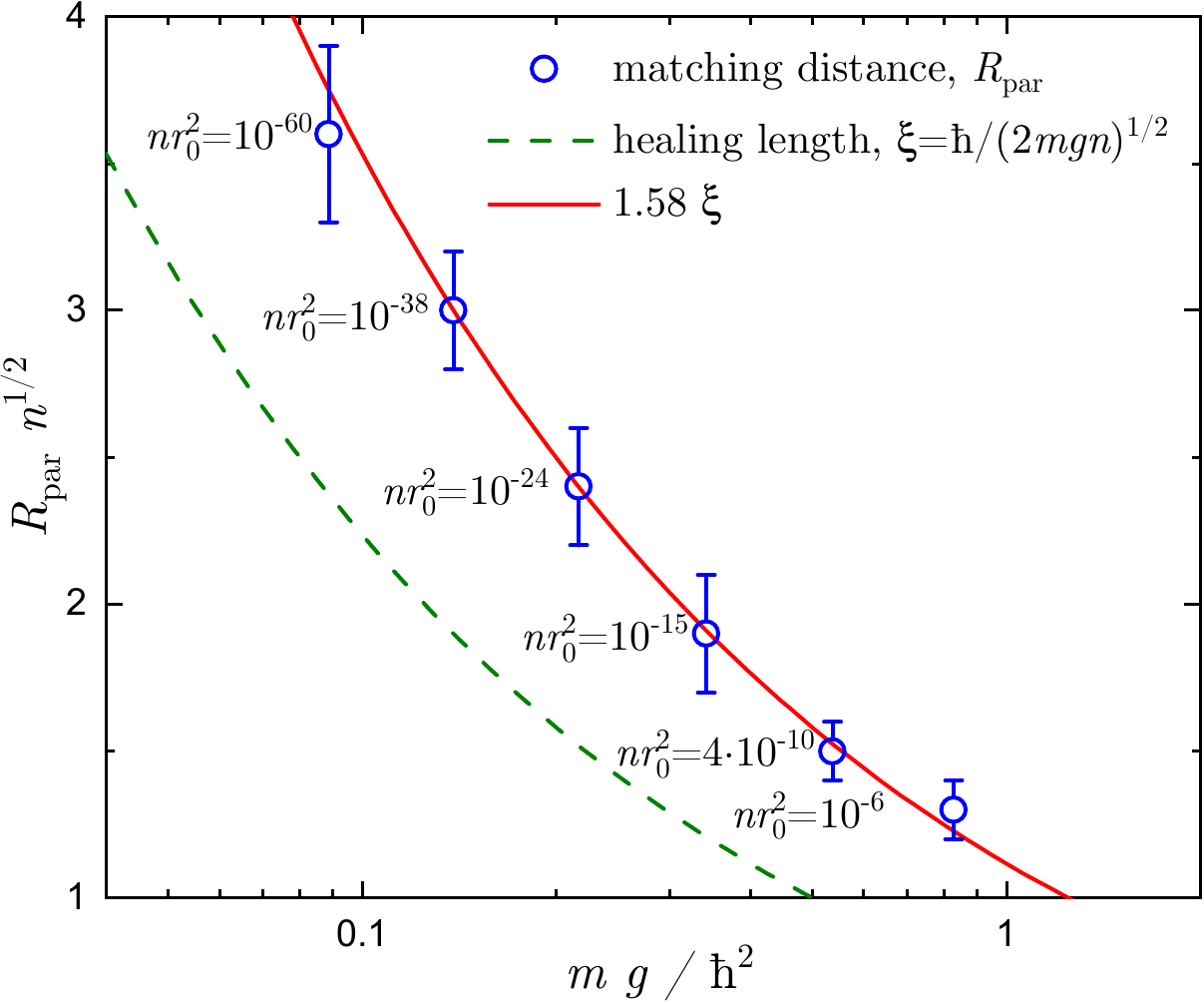}
\caption{Maximal distance up to which the two-body solution applies in the many-body problem in the rarefied limit.
Blue symbols: matching distance $R_{\rm par}$ between the two-body scattering solution and the hydrodynamic long-range asymptotic as obtained from variational Monte Carlo optimization of the energy.
Red solid line: the healing length $\xi$ scaled by a numerical constant $1.58 \xi$.
Distances shown on the vertical axis are given in the units of the mean interparticle distance $n^{-1/2}$.
The horizontal axis shows the 2D coupling constant.
}
\label{Fig_Rpar2D}
\end{figure}

Figure~\ref{Fig_Rpar2D} reports the optimal matching distance as a function of the energy per particle in the many-body system. We find that $R_{\rm par}$ scales linearly with the healing length over an extensive range of densities (with $nr_0^2$ ranging from $10^{-60}$ to $10^{-6}$). 
Although the considered densities might appear extremely low, a peculiarity of the two-dimensional geometry is that the coupling constant $g$, which is the input parameter in our simulations, depends on the density in a weak logarithmic way, so that the considered values of $g$ are not that small ($0.1 \lesssim g\lesssim 1$). To show that, we solve self-consistently the beyond-mean-field equation of state~(\ref{MoraCastin}) of a 2D Bose gas\cite{prl102180404}
\begin{equation}
\frac{mg}{\hbar^2}
=
\frac{4 \pi }{\ln \left(\frac{4 e^{-2-2 \gamma }}{na^2}
\frac{\hbar^2}{mg}
\right)}
\end{equation}
to extract the value of the coupling constant $g$ for each considered density $nr_0^2$. 
This suggests that the scattering problem is relevant up to distances on the order of the healing length, rather than the mean interparticle distance as suggested in previous studies~\cite{HofmannZwerger21}.

In order to demonstrate the excellent accuracy of the variational description, we compare the optimal variational energy (an upper bound) obtained using $R_{\rm par}$ shown in Fig.~\ref{Fig_Rpar2D} with the diffusion Monte Carlo energy (exact). We find a relative difference of only $10^{-4}$ for the coupling constant $g \approx 0.1 \hbar^2/m$. 
Such a tiny difference confirms that the variational wave function is very close to the true ground state, indicating that its properties closely reflect those of the ground state. Hence, the variational wave function can be used to define the region of the applicability of the two-body scattering solution in the many-body context.

While the hydrodynamic asymptotic in Eq.~\eqref{fr} is expected to fail at distances of the order of the healing length and smaller, we argue that, in variational calculations, the energy contributions from the hydrodynamic part (i.e. corresponding to the small momenta, $k\to 0$ of the linear excitation spectrum $v_s k$) are smaller compared to those corresponding to the short-distance (large momenta) behavior. Consequently, our variational procedure is primarily sensitive to the correct description of the short-range part, which is the focus of our validation in this study.

\section{Details of the integration}\label{app_int}
\subsection*{Two dimensions}
In order to calculate the pair correlation function~(\ref{g2_J0}) in the limit of small distances, $r\ll\xi$, we express it in terms of the healing length $\xi=\hbar/(\sqrt{2}mc)$ as $g_2(r)/n^2 = 1 + I/(\pi n\xi^2b^2)$, where 
\begin{equation}\label{int_devision}
  I = \int\limits_0^\infty dx J_0(x)\left(\frac{x^2}{\sqrt{x^2+b^2}} - x\right) = \int\limits_0^\infty dx J_0(x)\left(\sqrt{x^2 + b^2} - x\right) - b^2 \int\limits_0^\infty dx\frac{J_0(x)}{\sqrt{x^2 + b^2}}
\end{equation}
with $x = pr/\hbar$ and $b^2=2(r/\xi)^2$. Both integrals in the sum (\ref{int_devision}) will be considered as functions of $c$ in the limit $c\to0$. The second is a table integral
$$\mathcal{I}_2(b) \equiv \int\limits_0^\infty dx \frac{J_0(x)}{\sqrt{x^2 + b^2}} = I_0\left(\frac{b}{2}\right) K_0\left(\frac{b}{2}\right),$$
while the first
$$\mathcal{I}_1(b) \equiv \int\limits_0^\infty J_0(x)\left(\sqrt{x^2+b^2}-x\right)$$
can be found from the differential equation $\mathcal{I}_1^\prime(b) = b \mathcal{I}_2(b)$ with the initial condition $\mathcal{I}_1(0)=0$, so that
$$\mathcal{I}_1(b) = \int\limits_0^b dt I_0\left(\frac{t}{2}\right) K_0\left(\frac{t}{2}\right)t = \frac{b^2}{2}\left[I_0\left(\frac{b}{2}\right) K_0\left(\frac{b}{2}\right) - I_0^\prime\left(\frac{b}{2}\right)  K_0^\prime\left(\frac{b}{2}\right)\right].$$
Using the asymptotic behavior of the modified Bessel functions $I_0(b/2\to0) \approx 1 + b^2/16 + \mathcal{O}(b^3)$ and $K_0(b/2\to0)\approx \ln(4/be^{\gamma_E}) + b^2/16 \,\ln(4e/be^{\gamma_E}) + \mathcal{O}(b^3)$, we get
\begin{equation}
\lim\limits_{b\to0}\frac{g_2(r)}{n^2} = 1 - \frac{1}{\pi n \xi^2}\ln\frac{4}{be^{\gamma_E+1/2}} + \mathcal{O}(b^2) = 1 - \frac{m^2c^2}{2\pi\hbar^2n} \,\ln\frac{2\sqrt{2}}{e^{\gamma_E + 1/2}}\frac{\xi}{r} + \mathcal{O}\left(\frac{r^2}{\xi^2}\right).
\end{equation}

\subsection*{Three dimensions}
The integral in Eq.~(\ref{g2_3D_Bog}) is calculated likewise. Introducing the notation $16\pi anr^2 = \eta^2$, or $\eta = \sqrt{2} r/\xi \ll1$, we get
\begin{equation}
  \frac{g_2(r)}{n^2} = 1 - \frac{8a}{\pi r}\left[\int\limits_0^\infty\frac{\sin t}{\sqrt{t^2+\eta^2}}dt - \frac{1}{\eta^2}\int\limits_0^\infty (\sqrt{t^2+\eta^2} - t)\sin t dt\right].
\end{equation}
The first integral in the square brackets
$$\mathcal{I}_3(\eta) \equiv \int\limits_0^\infty\frac{\sin t}{\sqrt{t^2+\eta^2}}dt = \frac{\pi}{2}\bigl[I_0(\eta) - L_0(\eta)\bigr],$$
where $I_k(\eta)$ and $L_k(\eta)$ are the modified Bessel and Struve functions of $k$-th order, respectively. The second integral in the brackets
$$\mathcal{I}_4(\eta) \equiv \int\limits_0^\infty \sin t(\sqrt{t^2+\eta^2}-t)dt $$
satisfies $\mathcal{I}_4^\prime(\eta) = \eta\mathcal{I}_3(\eta)$ with the boundary condition $\mathcal{I}_4(0)=0$, which yields
$$\mathcal{I}_4(\eta) = \eta\frac{\pi}{2}\bigl[I_1(\eta)+L_1(\eta)\bigr].$$
As a result,
\begin{equation}\label{struve_ans}
\frac{g_2(r)}{n^2} = 1 - \frac{a}{r}\left[\frac{1}{\eta}L_1(\eta) - \frac{1}{\eta} I_1(\eta) + I_0(\eta) - L_0(\eta)\right]
\end{equation}
with the asymptotic at $\eta\to0$ given by Eq.~(\ref{g2_3D_final}). Note that in the asymptotic expression obtained from (\ref{struve_ans}), one first needs to take the limit $a\to 0$ and only then $r\to \infty$.

\section{Details of Monte Carlo simulations}\label{app_MC}

We perform diffusion Monte Carlo simulations to calculate the energetic and structural properties of a many-body system with power-law interactions. In three dimensions, we consider 
particle numbers $N = 10$, $13$, $17$, $25$, $50$, and $100$, which form an approximately linear sequence in $1/N$, 
convenient for performing the extrapolation to the thermodynamic limit. The energy per particle $E/N$ is obtained via a cubic fit in $1/N$ extrapolated to $1/N \to 0$. 

We calculate the ground state energy per particle $E/N$ on a density grid spanning a wide range, $na^3=10^{-5}$ to $1$, with each successive density twice the previous one.
The chemical potential is obtained from the density dependence of the energy per particle $E/N$, according to $\mu=\partial(n\; E/N)/\partial n$. 
The first derivative is evaluated using a 15-point central finite-difference formula. 
The derivative is evaluated at the midpoints of the logarithmic grid, effectively halving the grid spacing. 
We then interpolate the resulting chemical potentials onto the original density grid, so that both the energy per particle, $E/N$, and chemical potential $\mu$ are available at the same densities. Finally, the chemical potential is extrapolated to the thermodynamic limit using a quadratic fit in $1/N$, with the thermodynamic value obtained at $1/N=0$.

The contact $\mathcal{C}$ is determined from the short-distance behavior of the pair correlation function $g_2(r)$ using its short-range asymptotics [Eqs.~(\ref{g2_3D_asymp}) and (\ref{g2_2D}) in three and two dimensions, respectively], see Fig.~\ref{Figg2} for a characteristic example. 
Due to the divergent short-range repulsion, $g_2(r)$ is strongly suppressed and vanishes according to Eq.~(\ref{psi3D}) in 3D [and Eq.~\eqref{wavefunc2D} in 2D] geometry, corresponding to an effective hard-core behavior. In practice, values of $g(r)$ below the statistical noise are discarded, introducing a cutoff $R_{\min}$. The contact is extracted by linearly extrapolating $g(r)/\psi^2(r)$ to $r \to 0$ over the interval $R_{\min} < r < R_{\max}$ with $R_{\max}$ several times larger than $R_{\min}$.
We find that there is little difference between VMC and DMC results for the contact, so we use linear extrapolation. 
For Tan's contact, we use a linear fit in the inverse number of atoms $1/N$, in order to estimate its thermodynamic value $1/N=0$.


\end{document}